\documentstyle[12pt]{article}
\def\bbox{{\,\lower0.9pt\vbox{\hrule \hbox{\vrule height 0.2 cm

\hskip 0.2 cm

\vrule  height 0.2 cm}\hrule}\,}}

\def\bbox{{\,\lower0.9pt\vbox{\hrule \hbox{\vrule height 0.2 cm

\hskip 0.2 cm

\vrule  height 0.2 cm}\hrule}\,}}
\begin{document}
\setlength{\unitlength}{1mm}
\title{{\hfill {\small  } } \vspace*{2cm} \\
Kaluza-Klein Method in Theory of Rotating Quantum Fields}
\author{\\
D.V. Fursaev\thanks{e-mail: fursaev@thsun1.jinr.ru}
${}^{1}$
\date{}}
\maketitle
\noindent  
 \\ 
$^{1}${\em Joint Institute for Nuclear Research, Bogoliubov
Laboratory of Theoretical Physics, \\
141 980 Dubna, Russia} 
\bigskip

\begin{abstract}
Quantum fields on a stationary space-time
in a rotating Killing reference frame are considered.
Finding solutions of wave equations in this frame 
is reduced to a fiducial problem on a static 
background. The rotation results in a 
gauge connection in a way similar
to appearance of gauge fields in Kaluza-Klein models.
Such a Kaluza-Klein method in theory of rotating quantum fields
enables one to simplify computations and
get a number of new results similar to those established 
for static backgrounds.  
In particular, we find with its help functional form 
of free energy at high temperatures. 
Applications
of these results to quantum fields near rotating black holes
are briefly discussed.
\end{abstract}

\bigskip

{\it PACS number(s): O4.62.+v, 03.65.Pm, 04.70.Dy}

\bigskip

\baselineskip=.6cm

\noindent
\newpage
\section{Introduction}
\setcounter{equation}0

It is very well known that computations of quantum 
effects are simplified on static space-times where
some general results such as   
high-temperature asymptotics \cite{DK}--\cite{DSb}
can be established.  
Sometimes explicit computations can be done in case of
rigidly rotating fields if the background is
static and axially-symmetric. 
Recent results of this kind are 
high-temperature asymptotics 
of rotating CFT's on Einstein manifolds studied in     
\cite{HHT}--\cite{LL}.
Other similar computations can be found in 
\cite{CH}.  However, methods used for rotating fields
on static space-times cannot be applied to other important 
situations, to the Kerr geometry, for example.
The aim of this paper is to 
suggest an approach how to deal with 
quantum effects on stationary geometries  
and to get with its help some new results. 

\bigskip

The basic idea of our approach was formulated in 
\cite{FF:99a} and it is to reduce the problem on a stationary background
$\cal M$ to an equivalent but more simple
problem on a fiducial static space-time $\tilde{\cal M}$.  
This can be done as follows.
Consider a Killing frame of reference on $\cal M$, i.e., a
frame related to observers with
velocities $u^\mu$ parallel to the time-like Killing
field $\xi^\mu$. 
The space-time metric in this frame can be represented
as $ds^2=dl^2-B(u_\mu dx^\mu)^2$, where $B=-\xi^2$ and
$dl^2=h_{\mu\nu}dx^\mu dx^\nu$ 
is the proper distance between the points $x^\mu$ and 
$x^\mu+dx^\mu$. It can be shown that 
wave equations of fields on $\cal M$ are reduced to equations
on a fiducial static space-time $\tilde{\cal M}$ with metric
$d\tilde{s}^2=dl^2-Bdt^2$.
As a result of rotation of the frame the
covariant derivative on $\tilde{\cal M}$ takes the form
$\tilde{\nabla}_\mu-a_\mu \partial_t$, where 
$a_\mu dx^\mu=-u_idx^i/\sqrt{B}$.
For
a solution
$\phi_\omega(t,x^i)=e^{-i\omega t}\phi_\omega(x^i)$ 
with a certain frequency $\omega$
the vector $a_\mu$ is just a gauge
potential while $\omega$ is a charge. Appearance of $a_\mu$ in our 
case is analogous to appearance of the gauge potential in
Kaluza-Klein theories and for this
reason we call this method the Kaluza-Klein (KK) method.  
One can consider now a related static problem for a fiducial 
charged field $\phi^{(\lambda)}$ living on $\tilde{\cal M}$ 
and interacting with the gauge field $a_\mu$.
The charge of the field is $\lambda$, a real parameter.
If solutions $\phi^{(\lambda)}$  of field equations
on $\tilde{\cal M}$
are known for different $\lambda$ one can identify the single-particle
excitation of the physical field carrying energy $\omega$  
with the fiducial field having the same energy and charge 
$\lambda=\omega$, i.e., to put 
$\phi_\omega=\phi^{(\omega)}_\omega$.

In some cases the fiducial problem 
can be used and has certain
advantages. For instance, for fields near a rotating black hole
the spectrum of $\omega$ is continuous
and is specified by the density of levels $dn /d\omega$.
In this and other similar problems $dn/ d\omega$ plays an important
role in computing physical quantities, however,
direct derivation of $dn/d\omega$ by using eigen modes is
quite complicated.
On the other hand, for fiducial fields there are 
methods developed for
static backgrounds which do not require knowing eigen modes
explicitly.  
In this paper we establish relation between $dn/d\omega$
and the heat kernel of the operator $H^2(\lambda)$, where 
$H(\lambda)$ is the one-particle Hamiltonian of  
$\phi^{(\lambda)}$. We, thus, find a way how to compute 
$dn/d\omega$ by using powerful heat kernel techniques.  
This enables us to get a number of new
results.  
In particular, we compute the free of a rotating quantum
field at high temperatures 
\begin{equation}\label{i}
F(T)=-\int d^3x \sqrt{-g}\left[a T^4+T^2(\Phi+b\Omega^2)
+O(\ln T)\right]~~~.
\end{equation}
Here $a$, $b$ are numerical coefficients determined by the spin of  
fields, $T$ is the local Tolman temperature measured by the 
Killing observer.  The term $\Phi=\Phi(m,R,w)$ depends on the 
mass $m$ of the field, scalar
curvature $R$ and acceleration $w^\mu$ of the frame.  
This term and the coefficient $a$ are the same as for static spaces 
\cite{DSa}, \cite{DSb}.  The rotation results in the new term which 
depends on the angular velocity $\Omega$ of the Killing frame 
measured with 
respect to a local Lorentz frame.  In principle, our method also 
enables one to find explicitly other terms in temperature 
expansion (\ref{i}) and, in particular, $\ln T$ term in (\ref{i}) which 
we present later in the text.

The rest of the paper is organized as follows. 
The Kaluza-Klein method is
discussed in Section 2. We introduce
the Killing frame of reference, and
show how to  formulate the fiducial problem  
for scalar and spinor fields.
The systems with continuous spectrum are discussed in Section 3.
We introduce the density of levels $dn/d\omega$ and find out its 
relation to the heat kernel of $H^2(\lambda)$.  Applications of our 
method are discussed in the second part of the paper. By using the heat 
kernel technique we obtain high-frequency asymptotics of 
$dn/d\omega$ (Section 4) and with its help high-temperature 
behaviour of the free energy (Section 5).  
In Section 5 we also briefly discuss quantum theory near 
rotating black holes. Further comments and a discussion of
our results can be found in Section 6.
Some geometrical relations in the Killing frame are
presented in Appendix A.  
Appendix B clarifies 
some technical issues which appear in  Sections 3.

\section{The method}
\setcounter{equation}0

\subsection{Killing reference frame}

Let us consider a field $\phi$ on a domain $\cal M$ of a 
$D$-dimensional 
space-time where there is a time-like Killing vector field $\xi^\mu$
($\xi^2<0$).  
${\cal M}$ may be a complete  
manifold if 
$\xi^\mu$ is everywhere time-like. In most other cases $\xi^\mu$
is time-like only in some region. 
We will study solutions of field equations
in the frame of reference related to 
Killing observers
whose velocity $u^\mu$ 
is parallel to $\xi^\mu$
\begin{equation}\label{1a.1} 
u^\mu=B^{-1/2}\xi^\mu~~~,~~~B=-\xi^2~~~.
\end{equation}
For a Killing observer a solution $\phi_\omega$ 
carrying the energy $\omega$
is defined as 
\begin{equation}\label{1a.2}
i{\cal L}_\xi \phi_\omega=\omega \phi_\omega~~~
\end{equation}
where ${\cal L}_\xi$ is the Lie derivative along $\xi^\mu$.
The background metric
$g_{\mu\nu}$ can be represented as
\begin{equation}\label{1a.6}
g_{\mu\nu}=h_{\mu\nu}-u_\mu u_\nu~~~,
\end{equation}
where $h_{\mu\nu}$ is the projector on the directions
orthogonal to $u_\mu$.
Because sheer and expansion of the family of 
Killing trajectories
vanish identically the trajectories are characterized 
in each point only by their 
acceleration $w_\mu$ and the rotation 
$A_{\mu\nu}$ 
with respect to a local Lorentz frame \cite{HE}
\begin{equation}\label{1a.3}
w_\mu= u_{\mu;\lambda}u^\lambda~~~,
\end{equation}
\begin{equation}\label{1a.5}
A_{\mu\nu}=\frac 12 h^\lambda_\mu h^\rho_\nu
(u_{\lambda;\rho} - u_{\rho;\lambda})~~~.
\end{equation}
To proceed it is convenient to choose coordinates 
$x^\mu=(t,x^i)$, 
$i=1,D-1$, where $\xi=\partial/\partial t$ and, consequently,
$h_{0\mu}=0$.
According with 
(\ref{1a.1}), (\ref{1a.6}),
the interval on $\cal M$ can be written as
\begin{equation}\label{1a.7}
ds^2=-Bd\tau^2+dl^2~~~,
\end{equation}
\begin{equation}\label{1a.8}
d\tau=-{1 \over \sqrt{B}}(u_\mu dx^\mu)=dt+a_idx^i~~~,~~~
a_i=-{u_i \over \sqrt{B}}
\end{equation}
\begin{equation}\label{1a.9}
dl^2=h_{\mu\nu}dx^\mu dx^\nu=h_{ij} dx^i dx^j~~~.
\end{equation}
The vector $a_i$ can be used to synchronize the
clocks in points 
with coordinates $x^i$ and $x^i+dx^i$.
The metric $dl^2$
serves to measure the proper distance between these points.
In the coordinates $(t,x^i)$ the only non-zero components of
acceleration (\ref{1a.3})
and rotation (\ref{1a.5}) are
\begin{equation}\label{1a.10}
w_i=\frac 12 (\ln B)_{,i}~~~,~~~
A_{ij}= -\frac 12 \sqrt{B}(a_{i,j}-a_{j,i})~~~.
\end{equation}
In four-dimensional space-time one can define a vector of
local angular velocity
\begin{equation}\label{1a.11}
\Omega_i=\frac 12 \epsilon_{ijk} A^{jk}~~~,
\end{equation}
where $\epsilon_{ijk}$ is a totally antisymmetric tensor.
The absolute value 
of the angular velocity is 
\begin{equation}\label{1a.11b}
\Omega=(\Omega_i \Omega^i)^{1/2}=\left(\frac 12 A^{\mu\nu}
A_{\mu\nu}\right)^{1/2}~~~.
\end{equation}

The form of the metric in the Killing 
frame, equations (\ref{1a.7}), (\ref{1a.8}),
is preserved under arbitrary change of coordinates $x^i$
provided $h_{ij}$ and $a_i$ transform as a $D-1$ dimensional 
tensor and vector. There is also another group of
transformations, which preserves (\ref{1a.7}), (\ref{1a.8}),
namely, $t=t'+f(x)$, 
$a_i=a_i'-\partial_if(x)$, where $f$ is an arbitrary function of 
$x^i$.
Under these transformations
$a_i$ changes as an Abelian gauge vector field.
By considering single-particle excitations 
with the fixed energy 
$\omega$ 
\begin{equation}\label{4.1} 
\phi_{\omega}(t,x^i)=e^{-i\omega t}\phi_{\omega}(x^i)~~~,  
\end{equation} 
one can realize 
this group of transformations as a local $U(1)$,
$$
\phi_{\omega}(t,x^i)=
\phi'_{\omega}(t',x^i)=e^{-i\omega t'}\phi'_{\omega}(x^i)~~,~~  
$$
\begin{equation}\label{6.1} 
\phi_{\omega}(x^i)=
e^{i\omega f(x)}\phi'_{\omega}(x^i)~~~.  
\end{equation} 
In this picture, $\omega$ coincides with an "elementary 
charge". To quantize in the Killing frame one needs 
a full set of modes $\phi_{\omega}(x)$.  
As follows from the above 
arguments, the equations which determine 
$\phi_{\omega}(x)$ have a form of $D-1$ dimensional equations for 
charged fields in external 
gauge field $a_i$ 
on a space with the metric $h_{ij}$.
It is important that covariant properties
of the theory in
$D$ dimensions guarantee diffeo and 
gauge-covariant form of the $D-1$ dimensional problem.
Such a reduction from $D$ to $D-1$ 
is analogous to the Kaluza-Klein procedure which yields
the Einstein-Maxwell
theory from higher dimensional 
gravity. The difference between the two reductions is that in the 
standard Kaluza-Klein approach the "extra" dimensions are 
compact and the charges are quantized.

\bigskip

Let $\cal B$ denote a space with metric  $dl^2$, see
(\ref{1a.9}). 
If the Killing trajectories do not rotate,
$a_i=0$, $\cal B$ can be embedded in $\cal M$
as a constant-time hypersurface with the unit normal vector $u^\mu$. 
If $\Omega$ is not vanishing $\cal B$
cannot be embedded in $\cal M$ because
$u_\mu$ cannot be a gradient. 
Consider now a point $p$ on $\cal B$ with coordinates $x^i$ 
and a vector $V_i$ from the tangent space at $p$.  On $\cal M$, $p$ 
corresponds to a trajectory of a Killing observer with 
the same coordinates 
$x^i$.  At any point of the trajectory one can define a vector 
$V_\mu$ orthogonal to $u^\mu$ such as 
$V_i=h_i^\mu V_\mu$. 
Suppose that connection $\tilde{\nabla}_i$
on $\cal B$ is determined by $h_{ij}$. 
Then the covariant
derivative with respect to this connection can be written as 
\cite{HE}
\begin{equation}\label{A.4} 
\tilde{\nabla}_j V_i=
h^{\lambda}_i h^{\rho}_j V_{\lambda;\rho}~~~,
\end{equation}  
where $V_{\mu;\nu}$ is the covariant derivative on $\cal M$ with 
respect to the connection defined by $g_{\mu\nu}$.
(One can easily check that 
$\tilde{\nabla}_k h_{ij}=0$.)
Relation (\ref{A.4}) can be generalized to
an arbitrary field on $\cal M$.
For instance, for a scalar field 
\begin{equation}\label{4.7}
h_j^\mu\partial_\mu \phi=(\partial_j-a_j \partial_t)\phi
\equiv D_i\phi 
~~~,  
\end{equation}
for a vector orthogonal to $u^\mu$  
\begin{equation}\label{A.42} 
h^{\lambda}_i h^{\rho}_j V_{\lambda;\rho}
=(\tilde{\nabla}_j-a_j\partial_t) V_i
\equiv D_jV_i~~~,
\end{equation}  
where $V_i=h_i^\mu V_\mu$. The time derivative in 
(\ref{4.7}), (\ref{A.42}) appears in general
because fields on $\cal M$ change along the 
Killing trajectory.  If $\phi$ and 
$V_\mu$ are solutions 
with certain frequency, see (\ref{4.1}), then 
$D_i$ become covariant derivatives on $\cal B$ in 
external gauge field $a_i$. This demonstrates explicitly  
diffeo 
and gauge-covariance of the theory which are left after the reduction.

\subsection{Scalar fields}

To illustrate the KK method
we  consider first a real scalar field $\phi$ 
which satisfies the equation
\begin{equation}\label{4.3}
(-\nabla^\mu\nabla_\mu+V)\phi=0~~~,
\end{equation}
where $V$ is a potential.
In the Killing frame (\ref{1a.6})
the wave operator can be represented as
\begin{equation}\label{4.4}
\nabla^\mu\nabla_\mu=-{1 \over B}
(\xi^\mu\nabla_\mu)^2+{1 \over 2B} B^{,\nu}\nabla_\nu +
h^{\mu \nu}\nabla_\mu \nabla_\nu=
-{1 \over B}\partial_t^2-{1 \over 2B} B_{,i}h^{ij}D_j+
h^{ij}D_iD_j~~~,
\end{equation} 
where $D_i$ is defined by (\ref{4.7}), (\ref{A.42}).
It is easy to see that (\ref{4.4}) can be written
in a $D$--dimensional form
\begin{equation}\label{4.8}
\nabla^\mu\nabla_\mu=\tilde{g}^{\mu\nu}D_\mu D_\nu~~~,
\end{equation}
\begin{equation}\label{4.9}
D_\mu=\tilde{\nabla}_\mu-a_\mu \partial_t~~~,
\end{equation}
where $a_\mu dx^\mu=a_i dx^i$. The connections $\tilde{\nabla}_\mu$
are determined on some space $\tilde{\cal M}$
with the metric 
\begin{equation}\label{4.10}
d\tilde{s}^2=\tilde{g}_{\mu\nu}dx^\mu dx^\nu=-Bdt^2+dl^2~~~.
\end{equation}
Relation between $\tilde{\cal M}$ 
and $\cal M$ becomes transparent when comparing (\ref{4.10})
with (\ref{1a.7}). 
We will call 
$\tilde{\cal M}$ and $a_\mu$ the fiducial space-time and the fiducial
gauge potential, respectively. 
Let us consider now a scalar field 
$\phi^{(\lambda)}$ on $\tilde{\cal M}$ which obeys the equation
\begin{equation}\label{4.11}
(-\tilde{g}^{\mu\nu} (\tilde{\nabla}_\mu+i\lambda a_\mu) 
(\tilde{\nabla}_\nu+i\lambda a_\nu)+V)\phi^{(\lambda)}=0~~~.
\end{equation}
If $\phi^{(\lambda)}_\omega(t,x^i)
=e^{-i\omega t}\phi^{(\lambda)}_\omega(x^i)$
is a solution to (\ref{4.11}) then, as follows from
(\ref{4.8})--(\ref{4.11})
$\phi^{(\omega)}_\omega(t,x^i)$ is a solution
to (\ref{4.3}). Therefore, for a scalar filed 
the relativistic eigen-energy problem in the stationary
space-time can be reduced to an analogous problem on
a fiducial static background $\tilde{\cal M}$.
Equation (\ref{4.11}) can be further rewritten in the form
\begin{equation}\label{4.12}
H^2(\lambda)\phi^{(\lambda)}_{\omega}(x^i)=\omega^2 
\phi_{\omega}^{(\lambda)}(x^i)~~~,
\end{equation} 
where $H(\lambda)$ has the meaning of a relativistic single-particle
Hamiltonian for the field $\phi^{(\lambda)}_\omega(x^i)$
on $\cal B$. For definition and discussion of
single-particle Hamiltonians see \cite{FF:98a}.

\subsection{Spinor fields}

In the same way one can treat a free spin $1/2$ field $\psi$ 
described by the Dirac equation
\begin{equation}\label{Dirac}
(\gamma^\mu\nabla_\mu+m)\psi=0~~~.
\end{equation}
To this aim one has to represent the 
Dirac operator as 
\begin{equation}\label{4.13}
\gamma^\mu\nabla_\mu=\tilde{\gamma}^t \xi^\mu\nabla_\mu+
\tilde{\gamma}^iD_i~~~,
\end{equation}
where $\gamma_\mu$ and $\tilde{\gamma}_\mu$ are two sets of 
gamma-matrices on $\cal M$ and $\tilde{\cal M}$, respectively,
\begin{equation}\label{4.14}
\{\gamma_\mu,\gamma_\nu\}=2g_{\mu\nu}~~~,~~~
\{\tilde{\gamma}_\mu,\tilde{\gamma}_\nu\}=2\tilde{g}_{\mu\nu}~~~.
\end{equation}
\begin{equation}\label{4.15} 
\tilde{\gamma}_t=\xi^\mu\gamma_\mu~~~,~~~ 
\tilde{\gamma}_i=h_i^\mu \gamma_\mu~~~, 
\end{equation}
The spinor covariant derivatives are $\nabla_\mu=\partial_\mu+
\Gamma_\mu$ where $\Gamma_\mu$ are the connections.
By choosing the appropriate basis of one-forms (such that
$u_\mu dx^\mu$ is one of the elements of this basis) it is not
difficult to show that
\begin{equation}\label{4.16}
\xi^\mu \Gamma_\mu=\frac 14 \tilde{\gamma}^t \tilde{\gamma}^i B_{,i}=
\tilde{\Gamma}_t~~~,
\end{equation} 
where $\tilde{\Gamma}_t$ is the time-component of the spinor
connection on $\tilde{\cal M}$. 
The Dirac operator takes the form
\begin{equation}\label{4.17}
\gamma^\mu\nabla_\mu=\tilde{\gamma}^\mu(\tilde{\nabla}_\mu-a_\mu
\partial_t)=\tilde{\gamma}^\mu D_\mu~~~,
\end{equation}
where $\tilde{\nabla}_\mu=\partial_\mu+\tilde{\Gamma}_\mu$ 
are the spinor covariant derivatives on $\tilde{\cal M}$.
The corresponding equation and single-particle Hamiltonian
for fiducial spin 1/2 fields are
\begin{equation}\label{Dirac2}
(\tilde{\gamma}^\mu (\tilde{\nabla}_\mu+i\lambda a_\mu)
+m)\psi^{(\lambda)}=0~~~,
\end{equation}
\begin{equation}\label{Dirac3}
H(\lambda)=i\tilde{\gamma}_t
(\tilde{\gamma}^i (\tilde{\nabla}_i+i\lambda a_i)
+m)~~~.
\end{equation}
The solution to (\ref{Dirac2}) of the form
$\psi^{(\lambda)}_\omega(t,x^i)=
e^{-i\omega t}\psi^{(\lambda)}_\omega(x^i)$
solves (\ref{Dirac}) at $\lambda=\omega$.
We see, therefore, that the KK method is universal
for scalar and spinor fields in a sense that it does not depend
on field equations.  The fiducial background
$\tilde{\cal M}$ and the gauge field $a_\mu$ are determined
only  by the Killing vector and by geometry of
$\cal M$.

\subsection{Conformal transformation to 
zero acceleration space-time}

For practical purposes it is convenient to change representation 
of the single-particle Hamiltonians as \cite{FF:99a},\cite{FF:98a}
\begin{equation}\label{2.8}
\bar{H}(\lambda)=e^{-{D-k \over 2}\sigma} 
H(\lambda) e^{{D-k \over 2}\sigma}~~~, 
\end{equation} 
\begin{equation}\label{2.10}
e^{-2\sigma}=-\xi^2=B~~~,
\end{equation} 
where $k=2$ for scalars and $k=1$ for spinors.
$\bar{H}^2(\lambda)$ are second order differential
operators of the standard form
\begin{equation}\label{2.11}
\bar{H}^2(\lambda)=-\bar{h}^{ij}(\bar{\nabla}_i+i\lambda a_i)
(\bar{\nabla}_j+i\lambda a_j)+\bar{V}(\lambda)~~~.
\end{equation}
Connections $\bar{\nabla}_i$ 
correspond to fields on a $D-1$ space
$\bar{\cal B}$ conformally related to $\cal B$
\begin{equation}\label{2.12}
d\bar{l}^2=\bar{h}_{ij}dx^i dx^j=e^{2\sigma} dl^2~~~.
\end{equation}
For a scalar field described by (\ref{4.3}) the
"potential term" in (\ref{2.11}) is 
\begin{equation}\label{2.13}
\bar{V}(\lambda)=\bar{V}
=B\left[V+{D-2 \over 2}(\nabla^\mu w_\mu
-{D-2 \over 2} w^\mu w_\mu)\right]~~~,
\end{equation}
where $w_\mu$ is acceleration (\ref{1a.3}). 
For spinors
\begin{equation}\label{2.14}
\bar{V}(\lambda)=\frac 14 \bar{R}+
B(m^2+m\gamma^\mu w_\mu)-
i \sqrt{B}\lambda \gamma^\mu\gamma^\nu A_{\mu\nu}~~~, 
\end{equation}
where $m$ is the mass of the field, $\bar{R}$ is the scalar curvature 
of $\bar{\cal B}$ and $A_{\mu\nu}$ is the rotation tensor (\ref{1a.5}).
The relation between $\bar{R}$ and the scalar curvature $R$
of the physical space-time $\cal M$ is  (see Appendix A)
\begin{equation}\label{2.15}
\bar{R}=B[R+(D-1)(2\nabla^\mu w_\mu-(D-2)w^\mu w_\mu)-
A^{\mu\nu}A_{\mu\nu}]~~~.
\end{equation}
It is worth pointing out that
one can derive (\ref{2.11})  by a conformal
transformation in the initial equations on the
physical space-time. The physical metric $g_{\mu\nu}$ changes 
to $\bar{g}_{\mu\nu}=g_{\mu\nu}/B$ and the Killing vector 
on the rescaled space
has the unit norm, $\xi^2=-1$. Thus,
Killing observers in space-time $\bar{g}_{\mu\nu}$ have a 
non-vanishing angular velocity but zero acceleration.

\section{The density of energy levels}
\setcounter{equation}0

\subsection{Definition}

In what follows we will be dealing with problems
where spectrum of energies is continuous and show how the 
KK method can be used in this case. 
Such problems appear in a number of important physical
situations like quantum theory around rotating black holes.
It also turns out that computations in
case of continuous spectrum  are simplified.
A continuous spectrum is 
characterized by the density of levels which 
in non-relativistic quantum 
mechanics is defined as 
\begin{equation}\label{10.1} 
{dn(E)  \over dE}=\int d^3x \sum_{l} j_0(\Psi_{E,l})~~~.  
\end{equation} 
Here  
$\Psi_{E,l}$ is a complete set  of
eigen-functions of the energy operator with the same eigen-value $E$.
In (\ref{10.1}),  
$\sum_l j_0(\Psi_{E,l})$  
is the spectral density of states 
or the total spectral measure and $j_0$ is the time component
of the "current" 
\begin{equation}\label{10.2}
j_0(\Psi)=\Psi^{*} \Psi~~~,~~~
j_i(\Psi)=-{i \over 2m}\left(
\Psi^{*}\partial_i \Psi-
\partial_i\Psi^{*}\Psi \right)~~~,
\end{equation}
where $m$ is the mass of the particle.
The set of $\Psi_{E,l}$ is normalized
by using the delta-function $\delta(E-E')$ and, strictly speaking, 
the integral in r.h.s. of (\ref{10.1})
is divergent. To avoid the divergence one has to work with a 
regularized density obtained by restricting the
integration in (\ref{10.1}) to some compact region.

Similarly, in a relativistic quantum field theory one 
considers a
regularized density of energy
levels of single-particle excitations $\Psi_{\omega,l}$.
This quantity is defined by covariant 
generalization of (\ref{10.1}) as 
\begin{equation}\label{10.3} 
{dn(\omega)  \over d\omega}=
\int_{\Sigma} d\Sigma^\mu \sum_l j_\mu(\Psi_{\omega,l})~~~,  
\end{equation} 
where $\Sigma$ is a space-like Cauchy hypersurface and
$d\Sigma^\mu$ is its volume element.
The regularization means that 
$\Sigma$ is restricted by a region where integral (\ref{10.3})
converges.
The current $j_\mu$ is determined
by field equations and is divergence
free, $\nabla^\mu j_\mu=0$. This property guarantees independence
of (\ref{10.3}) on the choice on $\Sigma$.
For free scalar and spinor fields ($\Psi=\phi$ or $\psi$) 
\begin{equation}\label{10.4a}
j_\mu(\phi_1,\phi_2)=
-i(\phi^{*}_1 \partial_\mu\phi_2
-\partial_\mu\phi^{*}_1\phi_2)~~~,
\end{equation}
\begin{equation}\label{10.4b}
j_\mu(\psi_1,\psi_2)=
\bar{\psi}_1 \gamma_\mu\psi_2~~~,
\end{equation}
where $\bar{\psi}=\psi^{+}\tilde{\gamma}_t/\sqrt{B}$ is the
Dirac conjugate spinor.
The divergence of $j_\mu$ is zero if $\phi$ and $\psi$ obey 
equations (\ref{4.3}) and (\ref{Dirac}), respectively .
The density (\ref{10.3}) is 
obtained from (\ref{10.4a}), (\ref{10.4b}) when one takes
$\phi_1=\phi_2=\phi_{\omega,l}$ and
$\psi_1=\psi_2=\psi_{\omega,l}$.  The eigen modes
are assumed to be normalized with respect to
the products
\begin{equation}\label{10.5}
<\Psi_1,\Psi_2>\equiv\int_{\Sigma} 
d\Sigma^\mu j_\mu(\Psi_1,\Psi_2)~~~.
\end{equation}
In case of scalar fields (\ref{10.5}) is the 
standard Klein-Gordon product.

In the KK method
the single-particle modes are obtained as solutions
to a fiducial problem.
The
fiducial fields obey equations (\ref{4.11}) and
(\ref{Dirac2}) which dictate a different form of the 
corresponding vector currents and products, namely, 
\begin{equation}\label{10.6a}
\tilde{j}_\mu(\phi_1,\phi_2)=
-i(\phi^{*}_1 (\partial_\mu+i\lambda a_\mu)\phi_2
-(\partial_\mu-i\lambda a_\mu)\phi^{*}_1\phi_2)~~~,
\end{equation}
\begin{equation}\label{10.6b}
\tilde{j}_\mu(\psi_1,\psi_2)=
\bar{\psi}_1 \tilde{\gamma}_\mu\psi_2~~~,
\end{equation}
\begin{equation}\label{10.7}
(\Psi_1,\Psi_2)\equiv\int_{\tilde{\Sigma}} 
d\tilde{\Sigma}^\mu \tilde{j}_\mu(\Psi_1,\Psi_2)~~~.
\end{equation}
In general, (\ref{10.5}) and (\ref{10.7}) are
not equivalent and relation
between physical and fiducial modes requires 
an additional study.

\subsection{Scalar fields}

To find out this relation
we assume that the Killing frame
has zero acceleration and put $\xi^2=-1$.
This simplifies computations without loss of generality.
One can always reduce
a general problem to this case by  
a conformal rescaling, see Section 2.4.
Thus, on a constant-time hypersurface $\Sigma$
one has for (\ref{10.5}), (\ref{10.7})
\begin{equation}\label{10.8}
<\phi_\omega,\phi_\sigma>=\int_{\Sigma}\sqrt{h}d^{D-1}x
\left[(\omega+\sigma)
\phi^{*}_\omega \phi_{\sigma}+i\phi^{*}_\omega a^i(\nabla_i
+i\sigma a_i)\phi_\sigma-i(\nabla_i-i\omega a_i)\phi^{*}_\omega 
a^i\phi_\sigma \right]~~~,
\end{equation}
\begin{equation}\label{10.9}
(\phi^{(\lambda)}_\omega,\phi^{(\lambda)}_\sigma)=
\int_{\Sigma}\sqrt{h}d^{D-1}x
(\omega+\sigma)
(\phi^{(\lambda)}_\omega)^{*}\phi^{\lambda}_{\sigma}~~~,
\end{equation}
where $h=\det h_{ij}$, and $h_{ij}$ is the metric induced on $\Sigma$.
(Indexes $i,j$ are raised with the help of $h^{ij}$.)
In problems with a continuous spectrum $\Sigma$ 
has an infinite volume and
(\ref{10.8}), (\ref{10.9}) are to be 
interpreted in the sense of distributions. 
We denote $C_\infty$ the "asymptotic infinity" of $\Sigma$ where
integrals (\ref{10.8}), (\ref{10.9}) may diverge.
$C_\infty$ may be a true infinity of the space-time,
as in case of fields around a rotating star. 
However, in general, it is a region where the 
coordinate system associated with the given reference
frame is singular. 
For instance, $C_\infty$ may be a black hole horizon, or
a surface where rotation of the Killing frame approaches the 
speed of light.

The behaviour of fields at $C_\infty$ can be used
for normalization. This 
standard procedure 
is based on the fact
that inner products are reduced to surface integrals
over $C_\infty$. 
Let $C_r$ be a boundary of some finite region $\Sigma_r$
inside $\Sigma$ such that at $r\rightarrow\infty$ 
$\Sigma_r$ expands to $\Sigma$ and $C_r$ coincides with $C_\infty$.
Then, as is shown in Appendix B, at $\omega\neq\sigma$
\begin{equation}\label{10.24}
(\phi^{(\lambda)}_\omega,\phi^{(\lambda)}_\sigma)=
{1 \over (\omega-\sigma)}
\lim_{r\rightarrow \infty}\int_{C_r} d\sigma^i
\left(\nabla_i(\phi^{(\lambda)}_\omega)^{*}\phi^{(\lambda)}_\sigma-
(\phi^{(\lambda)}_\omega)^{*}\nabla_i\phi^{(\lambda)}_\sigma-
2i\lambda
a_i(\phi^{(\lambda)}_\omega)^{*}\phi^{(\lambda)}_\sigma
\right)~~,
\end{equation}
\begin{equation}\label{10.25}
<\phi_\omega,\phi_\sigma>=
{1 \over (\omega-\sigma)}
\lim_{r\rightarrow \infty}\int_{C_r} d\sigma^i
\left(\nabla_i\phi_\omega^{*}\phi_\sigma-
\phi_\omega^{*}\nabla_i\phi_\sigma-
i(\omega+\sigma)
a_i\phi_\omega^{*}\phi_\sigma\right)~~.
\end{equation}
Equations (\ref{10.24}) and (\ref{10.25}) are to be interpreted
in the sense of distributions.
Suppose now that 
$\phi^{(\lambda)}_\omega$ admit the following normalization
\begin{equation}\label{10.26}
(\phi^{(\lambda)}_{\omega,k},\phi^{(\lambda)}_{\sigma,l})=
\delta_{lk}\delta(\omega-\sigma)~~,
\end{equation}
where indexes $l,k$ correspond to additional degeneracy
of the wave-functions.
It is clear that if (\ref{10.26}) holds,
the modes $\phi_\omega=\phi^{(\omega)}_\omega$ have 
correct normalization 
with respect to the Klein-Gordon product, 
\begin{equation}\label{10.27}
<\phi_{\omega,k},\phi_{\sigma,l}>=
\delta_{lk}\delta(\omega-\sigma)~~~.
\end{equation}
This follows from the fact that  
at $\lambda=\omega$ (\ref{10.25})
is obtained from (\ref{10.24}) in the limit 
$\sigma \rightarrow \omega$. It is for this reason
identification $\phi_\omega$ with $\phi^{(\omega)}_\omega$  
is justified.

Let us consider now the regularized density of levels (\ref{10.3})
of single-particle states and show how it is related to 
the heat kernel 
of the operator $H^2(\lambda)$. 
Consider two integrals 
in the region $\Sigma_r$ 
\begin{equation}\label{10.18} 
{dn(\omega)  \over d\omega}=\int_{\Sigma_r} 
\sqrt{h}d^{D-1}x \sum_k
\left[2\omega
|\phi_{\omega,k}|^2+i\phi^{*}_{\omega,k} a^i(\nabla_i
+i\omega a_i)\phi_{\omega,k}-i(\nabla_i-i\omega a_i)
\phi^{*}_{\omega,k} 
a^i\phi_{\omega,k} \right]~~~,
\end{equation}
\begin{equation}\label{10.19}
{dn^{(\lambda)}(\omega)  \over d\omega}=
\int_{\Sigma_r}d\Sigma^\mu \sum_k
\tilde{j}_\mu (\phi^{(\lambda)}_{\omega,k})=
\int_{\Sigma_r} 
\sqrt{h}d^{D-1}x\sum_k
2\omega |\phi^{(\lambda)}_{\omega,k}|^2~~~.
\end{equation}
Quantity (\ref{10.19}) is the regularized spectral density of 
$H^2(\lambda)$.
Let us define also an auxiliary quantity
\begin{equation}\label{10.28}
{d\tilde{n}^{(\lambda)}(\omega)  \over d\omega}=
{dn^{(\lambda)}(\omega)  \over d\omega}
-{1 \over 4\lambda}\sum_k
\left[(\phi^{(\lambda)}_{\omega,k}, 
\partial_\lambda H^2(\lambda)\phi_{\omega,k}^{(\lambda)})
+(\phi^{(\lambda)}_{\omega,k}, 
\partial_\lambda H^2(\lambda)\phi_{\omega,k}^{(\lambda)})^{*}
\right]~~~,
\end{equation}
where according with (\ref{2.11}), (\ref{2.13}),
\begin{equation}\label{10.31}
\partial_\lambda H^2(\lambda)=-2ia^i(\nabla_i+i\lambda a_i)
-\nabla_i a^i~~~.
\end{equation}
As follows from (\ref{10.18}), (\ref{10.31}),
\begin{equation}\label{10.29}
{dn(\omega)  \over d\omega}=
{d\tilde{n}^{(\omega)}(\omega)  \over d\omega}~~~.
\end{equation}
Consider now the spectral representation for the
heat kernel of $H^2(\lambda)$
\begin{equation}\label{11.1}
\mbox{Tr}e^{-tH^2(\lambda)}=\int_{\mu}^\infty 
{dn^{(\lambda)}(\omega) \over d\omega} e^{-t\omega^2}d\omega~~,
\end{equation}
where integration in the trace
is restricted by $\Sigma_r$ .
The parameter $\mu$ ($\mu>0$)
is the mass gap
of $H^2(\lambda)$ 
and we assume that
there are no bound states in the spectrum. 
We also assume that
$\mu$ does not depend on $\lambda$,
which is true in a number of physical problems.
The spectral density can be written symbolically
as 
\begin{equation}\label{10.30}
{dn^{(\lambda)}(\omega)  \over d\omega}=
2\omega \mbox{Tr}[\delta(H^2(\lambda)-\omega^2)]~~~.
\end{equation}
One can also define the integral
\begin{equation}\label{10.32}
\int_{\mu}^\infty 
{d\tilde{n}^{(\lambda)}(\omega) \over d\omega} e^{-t\omega^2}d\omega
=\mbox{Tr}\left[\left(1-{1 \over 2\lambda}\partial_\lambda 
H^2(\lambda)\right)e^{-tH^2(\lambda)}\right]~~~,
\end{equation}
where the right hand side is the consequence of (\ref{10.28}).
Because the trace does not depend on the choice of the basis
and, hence, on $\lambda$ one can write (\ref{10.32}) as
\begin{equation}\label{10.33}
\int_{\mu}^\infty 
{d\tilde{n}^{(\lambda)}(\omega) \over d\omega} e^{-t\omega^2}d\omega
=\left(1 + {1 \over 2\lambda t}\partial_\lambda\right)
\mbox{Tr}e^{-tH^2(\lambda)}~~~. 
\end{equation}
This formula is our key relation which
together with (\ref{10.29}) enables us to
compute the physical density of levels $dn/d\omega$
by using powerful heat kernel techniques.
From (\ref{10.32}) one can also derive a formal
expression for $dn/d\omega$
$$
{dn(\omega)  \over d\omega}=
2\omega \mbox{Tr}\left[\left(1-{1 \over 2\lambda}\partial_\lambda 
H^2(\lambda)\right) \delta(H^2(\lambda)-\omega^2)
\right]_{\lambda=\omega}
$$
\begin{equation}\label{10.34}
=2\omega \mbox{Tr}\left[\delta(H^2(\lambda)-\omega^2) -{1 \over 
2\lambda} \partial_\lambda \theta(H^2(\lambda)-\omega^2) 
\right]_{\lambda=\omega}~, 
\end{equation} 
where $\theta(x)$ is the step function, and $\theta'(x)=\delta(x)$.
The above results concern
systems with continuous spectra. Some similar 
relations for discrete spectra are discussed in Appendix B.

\subsection{Spinor fields}

All results established for scalars can be 
extended to spin 1/2 fields. We again work on  
a zero-acceleration space-time.
Then, according to (\ref{10.4b}), 
(\ref{10.6b}) the products of spinor functions are
\begin{equation}\label{10.22}
<\psi_1,\psi_2>=\int_{\Sigma} \sqrt{h}dx^{D-1} \bar{\psi}_1
\gamma^t\psi_2~~~,~~~
(\psi_1,\psi_2)=\int_{\Sigma} \sqrt{h}dx^{D-1} \bar{\psi}_1
\tilde{\gamma}^t\psi_2 ~~~.
\end{equation}
These expressions are different because
$\gamma^t=
\tilde{\gamma}^t-a^i\tilde{\gamma}_i$.
They are reduced to the surface integrals
(see Appendix B)
\begin{equation}\label{10.35}
(\psi^{(\lambda)}_\omega,\psi^{(\lambda)}_\sigma)=
{i \over (\omega-\sigma)}
\lim_{r\rightarrow \infty}\int_{C_r} d\sigma^i
(\psi^{(\lambda)}_\omega)^{+}\tilde{\gamma}_i\psi^{(\lambda)}_\sigma~~,
\end{equation}
\begin{equation}\label{10.36}
<\psi_\omega,\psi_\sigma>=
{i \over (\omega-\sigma)}
\lim_{r\rightarrow \infty}\int_{C_r} d\sigma^i
\psi_\omega^{+}\tilde{\gamma}_i\psi_\sigma~~.
\end{equation}
Suppose that $\psi^{(\lambda)}_\omega$ are a set of 
modes properly normalized
in the sense of distributions, see (\ref{10.26}).
Then by comparing (\ref{10.35}), (\ref{10.36}) and 
using the same arguments as for scalar fields 
we conclude that modes 
$\psi_\omega=\psi^{(\omega)}_\omega$ are
normalized by (\ref{10.27}).
Let us introduce 
\begin{equation}\label{10.37}
{d\tilde{n}^{(\lambda)} \over d\omega}=
{d n^{(\lambda)} \over d\omega}-{\omega \over \lambda}
\sum_k(\psi^{(\lambda)}_{\omega,k},\partial_\lambda H(\lambda) 
\psi^{(\lambda)}_{\omega,k})~~~, 
\end{equation}
\begin{equation}\label{10.38}
\partial_\lambda H(\lambda)=-\tilde{\gamma}_ta^i\tilde{\gamma}_i
\end{equation}
where $H(\lambda)$ is the spinor Hamiltonian  (\ref{Dirac3})
and
\begin{equation}\label{10.39}
{d n^{(\lambda)} \over d\omega}=
\sum_k(\psi^{(\lambda)}_{\omega,k}, 
\psi^{(\lambda)}_{\omega,k})~~~ 
\end{equation}
is the spectral density of $H(\lambda)$.
Then, as follows from (\ref{10.22}), (\ref{10.37})--(\ref{10.39}),
the density of levels of physical states is
\begin{equation}\label{10.40}
{d n \over d\omega}=
\sum_k<\psi_{\omega,k},\psi_{\omega,k}>=\left.
{d\tilde{n}^{(\lambda)} \over d\omega}\right|_{\lambda=\omega}~~~.
\end{equation}
Finally, from (\ref{10.37}), (\ref{10.40}) one gets for
spinor density $dn/d\omega$ formula (\ref{10.33}).

\section{High-frequency asymptotics}
\setcounter{equation}0

Formula (\ref{10.33}) makes it possible to use
heat kernel techniques in stationary backgrounds 
and find $dn/d\omega$ is one important limit, namely,
in the limit of high frequencies $\omega$.
The integral in (\ref{10.33}) is 
determined by $\omega \simeq t^{-1}$
and this limit 
corresponds to the asymptotic form of the heat kernel at small 
values of $t$ 
\begin{equation}\label{2.4}
\mbox{Tr} e^{-t\bar{H}^2(\lambda)}
\simeq {1 \over (4\pi t)^{(D-1)/2}}
\sum_{n=0}^\infty\left[a_n(\lambda) t^n
+ b_n(\lambda) t^{n+\frac 12}\right]~~~,
\end{equation}
where $a_n$ and $b_n$
are the standard heat kernel coefficients, $n=0,1,2,...$.
On manifolds without boundaries $b_n=0$.
Coefficients $a_n(\lambda)$ and 
$b_n(\lambda)$ are even functions of $\lambda$ because
the fiducial theory is $U(1)$ invariant and the heat coefficients 
are even functions of charges.
The gauge invariance also guarantees that the
coefficients
are polynomials in powers of the Maxwell stress tensor and
its derivatives. In our case the role of the gauge field
is played by the vector $a_idx^i$, and hence the corresponding
Maxwell tensor is related to the rotation. In general,
\begin{equation}\label{2.16}
a_n(\lambda)=\sum_{m=0}^{[n/2]}\lambda^{2m}a_{2m,n}~~~,
~~~
b_n(\lambda)=\sum_{m=0}^{[n/2]}\lambda^{2m}b_{2m,n}~~~,
\end{equation}
where $a_{2m,n}$ do not depend on $\lambda$. 
The highest power of $\lambda$ in (\ref{2.16}) can be 
determined by analyzing dimensionalities.
Coefficients $a_0$ and $a_1$ in (\ref{2.4})
do not depend on $\lambda$.

The density of levels at high frequencies  can be found
from (\ref{2.4}) by using (\ref{10.33}).
In what follows, we assume that the mass gap of the operator
can be neglected. In this case
one can use the inverse Laplace transform in (\ref{10.33})
and simplify computations.
It should be noted, however, that for operators with
zero gap one has to take into account the presence of
infrared singularities which come out in (\ref{10.33}) at small
$\omega$. One of the possibilities to avoid this
problem  is to use the 
dimensional regularization and formally consider $D$ as a complex 
parameter.
It is instructive first to obtain the asymptotics for
the fiducial spectral density
\begin{equation}\label{2.5}
{dn^{(\lambda)} \over d\omega}\simeq 
{2 \omega^{D-2} \over (4\pi)^{(D-1)/2}}\sum_{n=0}^\infty\left[
{a_n(\lambda) 
\over \Gamma\left({D-1 \over 2}-n\right)} 
\omega^{-2n}+
{b_n (\lambda)
\over \Gamma\left({D-2 \over 2}-n\right)} 
\omega^{-(2n+1)}\right]~~~.
\end{equation} 
One can easily verify that for complex $D$ substitution
of (\ref{2.5}) in (\ref{11.1}) results in (\ref{2.4}).
For $d\tilde{n} /d\omega$ relation (\ref{10.33}) 
results in expansion of the same form
\begin{equation}\label{2.5a}
{d\tilde{n}^{(\lambda)} \over d\omega}\simeq 
{2 \omega^{D-2} \over (4\pi)^{(D-1)/2}}\sum_{n=0}^\infty\left[
{\tilde{a}_n(\lambda) 
\over \Gamma\left({D-1 \over 2}-n\right)} 
\omega^{-2n}+
{\tilde{b}_n (\lambda)
\over \Gamma\left({D-2 \over 2}-n\right)} 
\omega^{-(2n+1)}\right]~~~,
\end{equation} 
\begin{equation}\label{2.5b}
\tilde{a}_n(\lambda)=a_n(\lambda)+{1 \over 2\lambda}
\partial_\lambda a_{n+1}(\lambda)~~,~~ 
\tilde{b}_n(\lambda)=b_n(\lambda)+{1 \over 2\lambda}
\partial_\lambda b_{n+1}(\lambda)~~~. 
\end{equation}
Finally, by taking into account (\ref{10.29}), (\ref{2.16}), 
(\ref{2.5a}), (\ref{2.5b}) one finds the 
asymptotics of the physical density 
\begin{equation}\label{2.17}
{dn(\omega) \over d\omega}\simeq
{2 \omega^{D-2} \over (4\pi)^{(D-1)/2}}\sum_{n=0}^\infty\left[
{c_n
\over \Gamma\left({D-1 \over 2}-n\right)} 
\omega^{-2n}+
{d_n 
\over \Gamma\left({D-2 \over 2}-n\right)} 
\omega^{-(2n+1)}\right]~~~,
\end{equation}
\begin{equation}\label{2.18}
c_n=\sum_{m=n}^{2n}
{\Gamma\left({D-1 \over 2}-n\right)
 \over \Gamma\left({D-1 \over 2}-m\right)}
\left(a_{2(m-n),m}+(m-n+1)a_{2(m-n)+2,m+1}\right)~~~,
\end{equation}
\begin{equation}\label{2.19}
d_n=\sum_{m=n}^{2n}
{\Gamma\left({D-1 \over 2}-n\right)
 \over \Gamma\left({D-1 \over 2}-m\right)}
\left(b_{2(m-n),m}+(m-n+1)b_{2(m-n)+2,m+1}\right)~~~,
\end{equation}
where $\Gamma(x)$ is the gamma function.
It is remarkable that (\ref{2.17}) 
is a local functional expressed
in terms of the heat-kernel coefficients
of some differential operators. 

Some comments about (\ref{2.17}) are in order.

First, if there are no boundaries 
($b_n(\lambda)=0$) one gets from (\ref{2.17}) a finite result 
for even dimensions, although for odd $D$ the result
is formally zero.
As we will see in the next Section, the proper way of 
dealing with the infrared problem is to keep in 
(\ref{2.17}) $D$ complex till the last stage of computations.
Then both for even and odd $D$ the physical 
quantities determined with the help of $dn/d\omega$ 
are finite except, possibly,
a number of standard poles. 
As for $c_n$ and $d_n$, they remain finite for all $D$.

Second, it is interesting to note that 
when expansion in (\ref{2.17}) is approximated by
first two terms determined by $c_0$ and $c_1$
the physical and fiducial densities
coincide, $dn/d\omega\simeq 
dn^{(\omega)}/d\omega$. This
property can be helpful in computations, see \cite{FF:99a}. 

In general, the coefficients in the leading terms in
(\ref{2.17})
can be immediately computed by using (\ref{2.18}), (\ref{2.19}). 
The first coefficient is trivial, $c_0=a_0$.
According to 
(\ref{2.18}),
\begin{equation}\label{xx1}
c_1=a_1+\left({D-1 \over 2}-1\right)a_{2,2}~~~,
\end{equation} 
\begin{equation}\label{xx2}
c_2=a_2+\left({D-1 \over 2}-2\right)a_{2,3}+
\left({D-1 \over 2}-2\right)
\left({D-1 \over 2}-3\right)a_{4,4}~~,
\end{equation} 
where $a_n=a_{0,n}$. Hence, in four-dimensional space-time
\begin{equation}\label{5.6b}
c_1=a_1+\frac 12 a_{2,2}~~~,
\end{equation}
\begin{equation}\label{5.6c}
c_2=a_2-\frac 12 a_{2,3}+{3 \over 4}a_{4,4}~~.
\end{equation}
Term $a_{2,2}$ is determined by the
gauge part of $a_2(\lambda)$, see (\ref{2.16}), and
in four dimensions
\begin{equation}\label{11.2}
a_{2,2}=\alpha \int_{\Sigma_r}{\bar h}^{1/2} d^3x
\bar{F}^{ij}\bar{F}_{ij}~~~.
\end{equation}
where $\alpha=-1/12$ for scalars and $\alpha=r/6$ for spinors,
$r$ is the dimensionality of the spinor representation.
Expression (\ref{11.2}) can be rewritten in terms of  
local angular velocity (\ref{1a.11b}) if we note that
$\bar{F}_{ij}=F_{ij}=2A_{ik}/\sqrt{B}$ and
$\bar{F}_{ij}\bar{F}^{ij}=8B\Omega^2$. 

In order to compute $c_2$ one needs to know
contribution of gauge fields in $a_3(\lambda)$
and $a_4(\lambda)$. These terms for any spin
can be obtained from results of \cite{Avramidi},\cite{BGV}. 
For spin zero fields
$$ 
a_{2,3}= {1 \over 3!} \int_{\Sigma_r}{\bar 
h}^{1/2} d^3x \left[\frac 12 \left(\bar{V}-\frac 16\bar{R}\right) 
\bar{F}^{ij}\bar{F}_{ij}
+{1 \over 10}\bar{\nabla}^i \bar{F}_{ji}\bar{\nabla}_k \bar{F}^{jk}
\right.
$$
\begin{equation}\label{11.3}
\left.
-{1 \over 15}\bar{R}^{ij}\bar{F}_{ik}\bar{F}_j~^k
-{1 \over 30} \bar{R}^{ijkl}\bar{F}_{ij}\bar{F}_{kl}
\right]
~~~,
\end{equation}
\begin{equation}\label{11.4}
a_{4,4}={1 \over 4!}\int_{\Sigma_r}{\bar h}^{1/2} d^3x
\left[{1 \over 12} (\bar{F}^{ij}\bar{F}_{ij})^2
+{4 \over 15} \bar{F}_{ij}\bar{F}_{pk}\bar{F}^{ik}\bar{F}^{pj}
\right]~~~,
\end{equation}
where $\bar{V}$ is "potential term" (\ref{2.13}) of the
scalar operator $\bar{H}^2(\lambda)$, Eq. (\ref{2.11}).
For spinor fields
\begin{equation}\label{11.3b}
a_{2,3}=ra_{2,3}^{\mbox{\tiny{scal}}}-
{r \over 3!}\int_{\Sigma_r} \bar{h}^{1/2}d^3x
\left[{1 \over 12}(\bar{R}+12Bm^2) \bar{F}^{ij}\bar{F}_{ij}-
\frac 14 \bar{F}_{ij}\bar{\nabla}^2 \bar{F}^{ij}\right]~~,
\end{equation}
\begin{equation}\label{11.4b}
a_{4,4}=ra_{4,4}^{\mbox{\tiny{scal}}}+
{1 \over 4!}\int_{\Sigma_r} \bar{h}^{1/2}d^3x
\left[{1\over 16} \mbox{Tr}
(\bar{\gamma}^i\bar{\gamma}^j\bar{F}_{ij})^4-{r \over 2} 
(\bar{F}^{ij}\bar{F}_{ij})^2 \right]~~~,
\end{equation}
where $a_{2,3}^{\mbox{\tiny{scal}}}$ is given by (\ref{11.3}) with 
$\bar{V}=\frac 14 \bar{R}+Bm^2$ and 
$a_{4,4}^{\mbox{\tiny{scal}}}$ coincides with (\ref{11.4}).

Note that (\ref{11.2})--(\ref{11.4b}) are
expressed in terms of geometrical quantities of the
rescaled three-dimensional space with metric (\ref{2.12}).
All quantities can be also rewritten in terms
of the geometry of the physical space-time, acceleration 
and rotation of the chosen Killing reference frame.
For instance, by using 
(\ref{5.6b}), (\ref{11.2}), and (\ref{A.12})
one finds
that for scalar fields in $D=4$
\begin{equation}\label{11.5}
c_1=\int_{\Sigma_r}\sqrt{-g}d^3x {1 \over B}\left[\frac 16 
R-V-\frac 23 \Omega^2\right]~~~, 
\end{equation}
where $R$ is the scalar curvature of the physical space-time 
and $V$ is the scalar potential. For spinor fields
\begin{equation}\label{11.6}
c_1=r\int_{\Sigma_r}\sqrt{-g}d^3x {1 \over B}
\left[-{1 \over 12} R-\frac 12 (\nabla w-w^2)+
\frac 56 \Omega^2-m^2\right]~~~,
\end{equation}
where $\Omega$ is given in (\ref{1a.11b}).
Thus, rotation changes coefficients
starting with $c_1$.

\section{Some applications}
\setcounter{equation}0

\subsection{Vacuum energy}

Asymptotics (\ref{2.17}) can be used in a number
of applications. As a first example, consider 
computation of vacuum energy $E$
of a free quantum field on a stationary background 
\begin{equation}\label{14.1}
E=\int d\Sigma_\mu \xi_\nu \langle \hat{T}^{\mu\nu}\rangle_0~~~.
\end{equation}
Here $\hat{T}^{\mu\nu}$ is the stress-energy tensor
of the field,
$\xi$ is a time-like Killing vector, and the
integration goes over a space-like (Cauchy) hypersurface $\Sigma$.
It is convenient to choose $\Sigma$ as a constant time hypersurface. 
The quantum state is defined as a vacuum for single-particle 
excitations $\hat{\phi}_\omega$ with a certain energy,
${\cal L}_\xi \hat{\phi}_\omega=-i\omega \hat{\phi}_\omega$.
For a free field which  obeys equation (\ref{4.3}) 
the straightforward calculation gives the following
formal expression
\begin{equation}\label{14.2} 
E= \int_{\mu}^\infty 
d\omega \frac 12\omega {dn \over d\omega}~~~.
\end{equation}
At large $\omega$ the integral is divergent and 
one can use (\ref{2.17}) to study the form of the divergence.
In dimensional regularization  
the divergent
part of (\ref{14.2}) in four-dimensional theory is
\begin{equation}\label{14.3}
E_{\mbox{\tiny{div}}}=
{\mu^{D-4} \over (4\pi)^{D/2}} c_2{1 \over D-4}~~,
\end{equation}
where $c_2$ is determined by (\ref{5.6c}),
(\ref{11.2})--(\ref{11.4}). It would be interesting to
investigate relation of (\ref{14.3}) and 
corresponding divergence
of the vacuum energy computed by standard covariant methods.

\subsection{High-temperature asymptotics}

Let us consider a quantum state of fields on $\cal M$
which is viewed by a Killing observer as a thermal state
at the temperature $T=(\beta \sqrt{B})^{-1}$, where
$B=-\xi^2$ and $\beta$ is a positive constant.
Certainly, to ensure the thermal equilibrium there must exist
necessary physical conditions.
We assume that in systems we study these conditions are satisfied. 
In this case one can describe the system by a canonical ensemble
and introduce the free energy 
\begin{equation}\label{2.1}
F[\beta] = \eta\beta^{-1}\int_{\mu}^{\infty} d\omega {dn(\omega) \over 
d\omega} \ln (1-\eta e^{-\beta \omega})~~~, 
\end{equation} 
where $\eta=+1$ for bosons and $\eta=-1$ for fermions.  
At high temperatures
the parameter $\beta$ is small
and the dominant contribution in (\ref{2.1}) comes  out
from large frequencies $\omega\simeq \beta^{-1}$
where one can use the high-frequency asymptotics.
By using (\ref{2.17}) for $dn/d\omega$ in (\ref{2.1})
and by neglecting the gap 
one finds
\begin{equation}\label{2.20}
F(D,\beta)=
F_1(D,\beta)+
F_2(D,\beta)~~~,
\end{equation}
\begin{equation}\label{2.21}
F_1(D,\beta)=-
{1 \over \pi^{D/2}\beta^D}
\sum_{n=0}\gamma_{D,n}
\Gamma\left({D-2n \over 2}\right)
\zeta(D-2n)
c_n\left({\beta \over 2}\right)^{2n}
~~~,
\end{equation}
\begin{equation}\label{2.22}
F_2(D,\beta)=-
{1 \over \pi^{D/2}\beta^D}
\sum_{n=0}\gamma_{D-1,n}
\Gamma\left({D-2n-1 \over 2}\right)
\zeta(D-2n-1)
d_n\left({\beta \over 2}\right)^{2n+1}
~~~.
\end{equation}
Here $\zeta(x)$ is the Riemann zeta-function. The coefficient
$\gamma_D=1$ for bosons, and $\gamma_D=1-2^{2n+1-D}$ for fermions.
It should be noted that in case
of Bose fields (\ref{2.20}) includes also an additional contribution
${1 \over \beta}\int d\omega \ln (\beta\omega)dn/d\omega$
which appears at small $\omega$. 
Function $F_2$ is a pure boundary part of the free energy.
It follows from (\ref{2.21}) and (\ref{2.22}) that
$F_2$ is related to $F_1$
\begin{equation}\label{2.23}
F_2(D,\beta)
=
{1 \over \sqrt{4\pi}} 
\left. F_1(D-1,\beta)\right|_{c_n\rightarrow d_n}~~
\end{equation}
and it is sufficient to investigate $F_1$ only.

First, we remind that (\ref{2.21}) is obtained in 
dimensional regularization.  When parameter $D$ coincides with the 
physical dimensionality one of the terms in (\ref{2.21}) has a simple 
pole. This pole
corresponds to an infrared singularity of the theory with 
zero mass gap.  The pole in $F_1$ appears at $n=D/2$, for $D$ even and 
at $n=(D-1)/2$ for $D$ odd. 
By taking this into account one finds for (\ref{2.21}) in three
dimensions
$$ 
F_1(D=3,\beta)
\simeq -\gamma_{3,0}{\zeta(3) \over 2\pi} 
{c_0 \over \beta^3}-
\gamma_{3,1}
{c_1 \over 4\pi\beta}\left({1 \over D-3}-\ln(\beta\rho) \right)
$$ 
\begin{equation}\label{5.5a}
-{1 \over \pi^{3/2}}\sum_{n=2}\gamma_{3,n}
\Gamma\left(\frac 32-n\right)
\zeta(3-2n)c_n\left({\beta \over 2}\right)^{2n-3}~~,
\end{equation}
where $\rho$ is a dimensional parameter related to the regularization.
As we pointed out above, the free energy is not trivial, although 
density of levels (\ref{2.17}) used for its computations
vanishes if one goes to $D=3$. 
To get from (\ref{2.21}) the result  in four dimensions we use
the identity
$$
\Gamma(z/2)\zeta(z)=\pi^{z-1/2}\Gamma((1-z)/2)\zeta(1-z)~~~.
$$
It gives
$$ 
F_1(D=4,\beta)
\simeq -\gamma_{4,0}{\pi^2 \over 90} 
{c_0 \over \beta^4}-
\gamma_{4,1}
{1 \over 24} {c_1 \over \beta^2}+\gamma_{4,2}
{1 \over 16 \pi^2}\left({1 \over D-4}-\ln(\beta\rho) \right)c_2
$$ 
\begin{equation}\label{5.5}
-{1 \over 16\pi^{5/2}}\sum_{n=3}\gamma_{4,n}
\Gamma\left(n-\frac 32\right)
\zeta(2n-3)c_n\left({\beta \over 2\pi}\right)^{2n-4}~~~,
\end{equation}
When the frame does not rotates (\ref{5.5})
coincides with well known high temperature expansion 
\cite{DK}. 
Special interest is leading terms in (\ref{5.5}).
For scalar fields
\begin{equation}\label{12.1}
F_1(\beta)\simeq  
-\int d^3x \sqrt{-g}\left[{\pi^2 \over 90}  T^4
+{1 \over 24} T^2\left(\frac 16 R-V-\frac 23 \Omega^2
\right)+O(\ln T)\right]
\end{equation}
where $T$ is the local temperature.
For spinor fields
$$
F_1(\beta)\simeq  
-r\int d^3x \sqrt{-g}\left[{7\pi^2 \over 720}  T^4\right.
$$
\begin{equation}\label{12.3}
\left.
-{1 \over 48} T^2\left({1 \over 12} R+\frac 12 
(\nabla_\mu w^\mu-w_\mu 
w^\mu) - \frac 56 \Omega^2+m^2\right)+O(\ln T)\right]~. 
\end{equation} 
In the both cases rotation results in a new 
term $\sim T^2\Omega^2$.
In principle, our results enable one to compute next
terms in high-temperature expansion. In particular,
the logarithmic correction ($\ln T$)
to (\ref{12.1}) can be found explicitly with the 
help of (\ref{11.3})--(\ref{11.4b}).

\subsection{Quantum fields around rotating black holes}

One of the applications where asymptotics (\ref{12.1}) and (\ref{12.3})
can be used is studying a
quantum state of fields around a rotating 
black hole when fields are in thermal equilibrium and
rigidly rotate with black hole
with the same angular velocity $\Omega_H$. This
state is analogous to the Hartle-Hawking vacuum
known for Schwarzschild black holes. It was studied in \cite{FT} 
and recently discussed in \cite{OW1},\cite{OW2}.
To ensure thermal equilibrium between the black hole and fields
one has to surround the black hole by a reflecting mirror which 
has to rotate with the velocity $\Omega_H$. 

Consider a
Kerr-Newman black hole with the mass $M$, the charge $Q$,
and the angular momentum $J=aM$.
The metric in Boyer-Lindquist coordinates is
$$
ds^2=-\left(1-{2Mr-Q^2 \over \Sigma}\right)dt^2-
2{(2Mr-Q^2)a\sin^2\theta \over \Sigma}dtd\varphi
$$
\begin{equation}\label{215}
+{\Sigma \over \Delta} dr^2+\Sigma d\theta^2+{A \sin^2\theta
\over \Sigma}d\varphi^2~~~,
\end{equation}
\begin{equation}\label{216}
\Delta=r^2-2Mr+a^2+Q^2~~~,~~~
\Sigma=r^2+a^2\cos^2\theta~~~,
\end{equation}
\begin{equation}\label{217}
A=(r^2+a^2)^2-\Delta a^2\sin^2\theta~~~.
\end{equation}
The horizon is located at
\begin{equation}\label{218}
r=r_+=M+\sqrt{M^2-Q^2-a^2}~~~.
\end{equation}
The surface gravity $\kappa$ and the angular velocity
$\Omega_H$ for the Kerr-Newman black hole are
\begin{equation}\label{219}
\kappa={r_+-M \over r_+^2+a^2}~~~,~~~\Omega_H={a \over r_+^2+a^2}~~~.
\end{equation}
Fields in thermal
equilibrium with the black hole are described by 
a canonical ensemble in the Killing frame with the Killing vector
$\xi=\partial_t+\Omega_H\partial_\varphi$. 
The local temperature is
$T=\kappa/(2\pi\sqrt{B})$ where $B=-g_{tt}-2\Omega_Hg_{t\varphi}
-\Omega_H^2g_{\varphi\varphi}$ and $g_{\mu\nu}$ are defined in
(\ref{215}). On the horizon $B=0$.
Thus, near the horizon the local temperature is large
and asymptotics (\ref{12.1}), (\ref{12.3}) are very good 
approximation
for the free energy. The result can be easily found by
using formula $w_\mu=\nabla_\mu \ln B/2$ for acceleration
and  formula (\ref{1a.11b}) for angular velocity.
The fiducial gauge potential which determines $\Omega$ has
only one non-zero component 
$a_\varphi=-(g_{t\varphi}+\Omega_Hg_{\varphi\varphi})/B$.

In principle, because (\ref{12.1}) and (\ref{12.3}) are functionals
of an arbitrary metric these expressions can be used to extract
more information about the quantum state. If the
free-energy is considered as
a thermal part of quantum effective action then
(\ref{12.1}) and (\ref{12.3}) can be used to derive the
stress energy tensor. 
Note that (\ref{12.1}), (\ref{12.3}) admit stationary variations
$\delta g_{\mu\nu}$ of the metric 
(${\cal L}_\xi \delta g_{\mu\nu}=0$)
when components $\xi^\mu$ of the Killing vector are held fixed. 
By considering such a variation of the first leading term
in (\ref{12.1}) one finds for scalar fields
\begin{equation}\label{stress}
<T^{\mu\nu}>_T=-{2 \over \sqrt{-g}}{\delta F_1 \over \delta g_{\mu\nu}}
={\pi^2 \over 90} T^4\left(g^{\mu\nu}-4{\xi^\mu \xi^\nu \over \xi^2}
\right)~~~.
\end{equation} 
Tensor $<T^{\mu\nu}>_T$ 
is divergence free and traceless and it corresponds
to the stress tensor of thermal radiation around a black hole.
It diverges on the black hole horizon where $T$ is infinite.
It also diverges at the surface
where the Killing frame rotates with the velocity of light
in agreement with arguments of \cite{FT}.
However in this case our results cannot be much trusted.
By using (\ref{12.1}) and similar variational procedure
one can find corrections to (\ref{stress}) due to curvature,
acceleration and rotation.  We are planing to study 
$<T^{\mu\nu}>_T$ in a separate publication.

\section{Summary and comments}
\setcounter{equation}0

The aim of our paper was to develop a computation method
applicable to rotating quantum fields and 
to get with its help new general results.
We have shown, in particular, that asymptotic form
of free energy at high temperatures 
can be found in terms of the heat kernel coefficients
of some differential operators. These operators
are interpreted as one-particle Hamiltonians of a
fiducial problem in external Abelian gauge field on a 
static background. We hope that in some cases 
where computations are quite involved, 
like rotating black holes,
our method will be the helpful and effective tool.
It would be an interesting problem to 
compare our method with covariant Euclidean formulation
of finite-temperature theory.

We considered here scalar and spinor fields.
Spin 1 fields require additional study to resolve a technical 
difficulty connected with constraints.  

Our analysis was restricted by systems with 
continuous spectrum. An advantage of a continuous spectrum
is that it is specified by the density of 
levels and we are able to relate 
the latter to the heat kernel of fiducial 
Hamiltonians.  The disadvantage is that one has to
work with regularized quantities. 
In \cite{BP}--\cite{LL} high-temperature asymptotics 
of rotating fields were obtained on Einstein manifolds
where the space is $S^3$. In this case operators have discrete spectra,
which can be found explicitly for conformal fields.
A naive calculation of our asymptotics (\ref{12.1}), (\ref{12.3})
on the Einstein manifold coincides with \cite{BP}--\cite{LL}
in the leading term proportional to $T^4$. 
However, the next, $T^2\Omega^2$ term, does not 
reproduce the result of \cite{BP},\cite{LL} and the discrepancy is not 
in numerical coefficients.  
Thus, it is an open question how (\ref{12.1}), (\ref{12.3})
are modified by finite-size effects.

\vspace{12pt}
{\bf Acknowledgements}:\ \ 
I am grateful to V.P. Frolov and D.V. Vassilevich for helpful 
discussions.
This work is supported in part by the RFBR grant
N 99-02-18146 and NATO Collaborative Linkage Grant, 
CLG.976417.

\newpage
\appendix
\section{Geometry in the Killing frame}
\setcounter{equation}0

We consider the metric $g_{\mu\nu}=h_{\mu\nu}-u_\mu u_\nu$ 
in the Killing frame characterized by the four-velocity $u_\mu$.  
In coordinates where $u^\mu=(1/\sqrt{B},0,0,0)$
the metric can be written in the form
\begin{equation}\label{A.1}
ds^2=-B(dt+a_idx^i)^2+h_{ij}dx^idx^j~~,
\end{equation} 
where $a_i=-u_i/\sqrt{B}$.
Metric tensor has the following components
\begin{equation}\label{A.2}
g_{\mu\nu}=\left(\begin{array}{cc}
h_{ij}-Ba_ia_j & -Ba_i \\
-Ba_j & -B
\end{array}\right)~~,~~
g^{\mu\nu}=\left(\begin{array}{cc}
h^{ij} & -a^i \\
-a^j & a^2-B^{-1}
\end{array}\right)~~~,
\end{equation}
where $h^{ij}h_{jk}=\delta^i_k$, $a^i=h^{ij}a_j$, $a^2=a^ia_i$.
As follows from (\ref{A.2}), $\det g_{\mu\nu}=-B\det h_{ij}$.
Relation (\ref{A.4}) enables one
to connect the Rieman tensors on $\cal M$
and $\cal B$
\begin{equation}\label{A.5}
R^\lambda_{~\mu\nu\rho}[h]=
R^\gamma_{~\alpha\beta\sigma}[g]h^\alpha_\mu h^\beta_\nu
h^\sigma_\rho h^\lambda_\gamma-A^\lambda_{~\nu}A_{\mu\rho}
+A^\lambda_{~\rho}A_{\mu\nu}-2A^{\lambda}_{~\mu}A_{\nu\rho}~~~,
\end{equation} 
where $A_{\mu\nu}$ is the rotation tensor (\ref{1a.5}).
Our conventions are 
$R^\sigma_{~\mu\nu\lambda}=\Gamma^\sigma_{\mu\lambda,\nu}-...$.
Formula (\ref{A.5}) is similar to embedding formula
for the Riemann tensor of a hypersurface, see \cite{HE}. 
The relation between the scalar curvatures
of $\cal B$ and $\cal M$, which follows from (\ref{A.5}) is
\begin{equation}\label{A.7}
R[h]=R[g]+2R_{\mu\nu}[g]u^\mu u^\nu-3 A^{\mu\nu}A_{\mu\nu}~~~.
\end{equation}
By taking into account that for the Killing field $\xi_\mu$
\begin{equation}\label{A.8}
\nabla^2\xi_{\mu}=-R^\lambda_\mu[g]\xi_\lambda~~~,
\end{equation}
we find with the help of  (\ref{1a.3}), (\ref{1a.5})
\begin{equation}\label{A.9}
R_{\mu\nu}[g]u^\mu u^\nu=\nabla_\mu w^\mu+A^{\mu\nu}A_{\mu\nu}~~~,
\end{equation}
where $w_\mu$ 
is the acceleration of the frame. Therefore,
\begin{equation}\label{A.10}
R[h]=R[g]+2\nabla_\mu w^\mu-A^{\mu\nu}A_{\mu\nu}~~~.
\end{equation}
In this paper we also introduced the space
$\bar{\cal B}$ which is conformally related to $\cal B$ 
\begin{equation}\label{A.11}
d\bar{l}^2=\bar{h}_{ij}dx^idx^j=
B^{-1}h_{ij}dx^idx^j~~~.
\end{equation} 
By using (\ref{A.5}) one can express geometrical quantities
on $\bar{\cal B}$ in terms of quantities on $\cal M$. In particular,
in four dimensions $D=4$, one has
$$
R[\bar{h}]=B\left(R[h]+4\tilde{\nabla}_iw^i
-2w^i w_i\right)
$$
\begin{equation}\label{A.12}
=B\left(R+6\nabla_\mu w^\mu-6w_\mu 
w^\mu-A^{\mu\nu}A_{\mu\nu}\right)~~~, 
\end{equation} 
where $\tilde{\nabla}_i$ is the connection on $\cal B$.
Finally, one can 
find relation  between $\cal M$ and fiducial space $\tilde{\cal M}$ 
with metric \begin{equation}\label{A.13} 
d\tilde{s}^2=-Bdt^2+h_{ij}dx^idx^j~~.
\end{equation} 
To this aim one should note that $\cal B$ can be embedded in 
$\tilde{\cal M}$ as a constant time hypersurface and find
embedding relation analogous to (\ref{A.5}) between 
$R^\lambda_{~\mu\nu\rho}[h]$ and
$R^\lambda_{~\mu\nu\rho}[\tilde{g}]$. By acting in this way 
we easily get
\begin{equation}\label{A.14}
R[\tilde{g}]=R[g]-A^{\mu\nu}A_{\mu\nu}~~~,
\end{equation}
and other similar identities.

\section{The inner products}
\setcounter{equation}0

Here we derive relations (\ref{10.24}), (\ref{10.25}), (\ref{10.35}),
and (\ref{10.36})
for inner products. 
For scalar functions one has
$$
(\omega^2-\sigma^2)(\phi^{(\lambda)}_\omega,\phi^{(\lambda)}_\sigma)=
(\phi^{(\lambda)}_\sigma,H^2(\lambda)\phi^{(\lambda)}_\omega)^{*}-
(\phi^{(\lambda)}_\omega,H^2(\lambda)\phi^{(\lambda)}_\sigma)=
$$
\begin{equation}\label{B.1}
(\omega+\sigma)
\lim_{r\rightarrow \infty}\int_{C_r} d\sigma^i
\left(\nabla_i(\phi^{(\lambda)}_\omega)^{*}\phi^{(\lambda)}_\sigma-
(\phi^{(\lambda)}_\omega)^{*}\nabla_i\phi^{(\lambda)}_\sigma-
2i\lambda
a_i(\phi^{(\lambda)}_\omega)^{*}\phi^{(\lambda)}_\sigma
\right)~~.
\end{equation}
This gives (\ref{10.24}). 
Note that (\ref{B.1}) also holds when 
$\Sigma$ has additional boundaries other than $C_\infty$
provided if fields obey Dirichlet conditions at these boundaries. 
To find (\ref{10.25}) we begin with
relation
\begin{equation}\label{B.2}
<\phi_\omega,\phi_\sigma>=(\phi_\omega,\phi_\sigma)+
\int_{\Sigma_r} \sqrt{h}d^{D-1}\left(i\phi_\omega^{*}a^i
(\nabla_i+i\sigma a_i)\phi_\sigma-ia^i(\nabla_i-i\omega a_i)
\phi_\omega^{*}\phi_\sigma\right)
\end{equation}
and use the identity
\begin{equation}\label{B.3}
H^2(\omega)-H^2(\sigma)=(\omega-\sigma)(-2ia^i\nabla_i-
i\nabla_i a^i+ (\omega+\sigma)a^ia_i)~~~.
\end{equation}
This enables one to rewrite (\ref{B.2}) as
$$
<\phi_\omega,\phi_\sigma>=(\phi_\omega,\phi_\sigma)
$$
\begin{equation}\label{B.4}
+{1 \over 2(\sigma^2-\omega^2)}\left[
(\phi_\sigma, (H^2(\omega)-H^2(\sigma))\phi_\omega)^{*}+
(\phi_\omega, (H^2(\omega)-H^2(\sigma))\phi_\sigma)\right]~~~.
\end{equation}
The right hand side of (\ref{B.4}) is a pure surface term
over $C_r$ which coincides with the right hand side of equation
(\ref{10.25}).

To prove (\ref{10.35}) for spinor modes it is sufficient to see that
\begin{equation}\label{B.5}
(\psi^{(\lambda)}_\omega,\psi^{(\lambda)}_\sigma)=
{1 \over (\omega-\sigma)}\left[
(\psi^{(\lambda)}_\sigma,H(\lambda)\psi^{(\lambda)}_\omega)^{*}-
(\psi^{(\lambda)}_\omega,H(\lambda)\psi^{(\lambda)}_\sigma)
\right]~~~
\end{equation}
and use the fact that $\bar{\psi}=\psi^{+}\tilde{\gamma}_t$ on
zero-acceleration space-time, see Section 2.4.
To prove (\ref{10.36})  we first note  that for spinor 
Hamiltonians
\begin{equation}\label{B.6}
H(\omega)-H(\lambda)=(\sigma-\omega)\tilde{\gamma}^t\tilde{\gamma}^ia_i~~~,
\end{equation}
hence
\begin{equation}\label{B.7}
<\psi_\omega,\psi_\sigma>=
(\psi_\omega,\psi_\sigma)+
{1 \over (\sigma-\omega)}\int_{\Sigma_r}\sqrt{h}d^{D-1}x
\psi_\omega^{+}(H(\omega)-H(\sigma))\psi_\sigma~~~.
\end{equation}
The right hand side of this equation coincides with the 
surface term in the right hand side of (\ref{10.36}).

In this Appendix we also comment on some properties
of operators with discrete spectrum.
Consider scalar operator $H^2(\lambda)$ which 
has a discrete spectrum 
$\omega^2(\lambda)$
\begin{equation}\label{C.1}
H^2(\lambda)\phi_\omega^{(\lambda)}=
\omega^2(\lambda)\phi_\omega^{(\lambda)}~~~.
\end{equation}
By differentiating the both sides of (\ref{C.1}) over $\lambda$
one finds
\begin{equation}\label{C.2}
\partial_\lambda 
H^2(\lambda)\phi^{(\lambda)}_\omega+H^2(\lambda)
\partial_\lambda\phi^{(\lambda)}_\omega= 
\partial_\lambda\omega^2(\lambda)\phi_\omega^{(\lambda)}
+\omega^2(\lambda)\partial_\lambda\phi_\omega^{(\lambda)}~~~,
\end{equation}
\begin{equation}\label{C.3}
\partial_\lambda H^2(\lambda)=-2ia^i(\nabla_i+i\lambda a_i)
-\nabla^i a_i~~~.
\end{equation}
The fiducial modes 
$\phi^{(\lambda)}_\omega$ now have a finite norm 
$(\phi^{(\lambda)}_\omega,\phi^{(\lambda)}_\omega)$
which we denote as $N^2(\lambda,\omega)$. 
Let $N^2(\omega)$ be the norm of  
physical functions $<\phi_\omega,\phi_\omega>$ .
The relation between the two norms follows from
definitions (\ref{10.8}), (\ref{10.9})
$$
N^2(\omega)=N^2(\omega,\omega)+\int_{\Sigma}
\sqrt{h}d^{D-1}x i\phi_\omega^{(*)}\left[2a^i(\partial_i+i\omega a_i)
+\nabla^i a_i\right]\phi_\omega=
$$
\begin{equation}\label{C.4}
=\left.
\left(1-{1 \over 2\lambda}\partial_\lambda \omega^2(\lambda)\right)
N^2(\omega,\lambda)\right|_{\lambda=\omega}~~~,
\end{equation}
where to get the last line we used (\ref{C.2}) and assumed that 
possible surface terms
which appear under integration by parts vanish 
due to boundary conditions.  Equation (\ref{C.4}) is an 
analog of equation (\ref{10.28}) obtained for continuous spectra.

\end{document}